\documentclass[pra,twocolumn,showpacs,amsmath,amssymb,superscriptaddress,floatfix]{revtex4-2}

\UseRawInputEncoding
\usepackage{amsmath}
\usepackage{xcolor}
\usepackage{graphicx}
\usepackage{float}
\usepackage{balance}
\usepackage{bbm}
\usepackage[colorlinks=true,allcolors=blue]{hyperref}

\begin{document}


\title{Recursive contact tracing in Reed-Frost epidemic models}


\author{Saumya Shivam}
\affiliation{Department of Physics, Princeton University, Princeton, New Jersey 08544, USA}
\author{Vir B. Bulchandani}
\affiliation{Department of Physics, University of California, Berkeley, Berkeley CA 94720, USA}
\affiliation{Princeton Center for Theoretical Science,  Princeton University, Princeton, New Jersey 08544, USA}

\author{S. L. Sondhi}
\affiliation{Department of Physics, Princeton University, Princeton, New Jersey 08544, USA}


\date{\today}

\begin{abstract}

We introduce a Reed-Frost epidemic model with recursive contact tracing and asymptomatic transmission. This generalizes the branching-process model introduced by the authors in a previous work [arxiv:2004.07237] to finite populations and general contact networks. We simulate the model numerically for two representative examples, the complete graph and the square lattice. On both networks, we observe clear signatures of a contact-tracing phase transition from an ``epidemic phase'' to an ``immune phase'' as contact-network coverage is increased. We verify that away from the singular line of perfect tracing, the finite-size scaling of the contact-tracing phase transition on each network lies in the corresponding percolation universality class. Finally, we use the model to quantify the efficacy of recursive contact-tracing in regimes where epidemic spread is not contained.
\end{abstract}


\maketitle

\section{Introduction}

One reason that the spread of COVID-19 has been difficult to contain using established methods of epidemic control is the high rate of viral transmission by pre-symptomatic and asymptomatic individuals~\cite{jhu_intro,wei2020presymptom,asymp_review,garcia2020occurence}. Although widespread vaccination is within reach at the time of writing~\cite{astrazeneca,moderna,pfizer}, new contagious diseases with a substantial rate of non-symptomatic transmission continue to pose a threat to global public health infrastructure.

Confronted with such a disease, the options available to policy makers are limited. One crude, effective and widely adopted intervention is restricting populations' movements via social distancing measures. A more efficient approach is the established technique of contact tracing~\cite{Fraser}, whereby the contacts of newly detected infected individuals are traced and isolated, ideally before they become contagious. However, for COVID-19, the rapid timescale on which an infected individual becomes contagious (on the order of a few days) can render traditional methods of manual contact tracing, with their attendant delays, completely ineffective~\cite{Fraser,Eames,ferretti2020quantifying}. It was realized early in the course of the COVID-19 epidemic that digital, app-based contact tracing might overcome these difficulties~\cite{ferretti2020quantifying,faggian2020proximity,Salathe,Yoneki}.

In a previous work~\cite{DHI}, we introduced the notion of ``digital herd immunity" as a precise way to quantify the efficacy of digital contact tracing. By applying ideas from percolation theory and the statistical physics of epidemic spread~\cite{Fisher,Cardy_1985} to contact tracing, we argued that successful digital contact-tracing protocols can be understood in terms of a ``contact-tracing phase transition'' to a collective, immune phase as the take-up of digital contact-tracing apps is increased. Our main finding was that regardless of the fraction of non-symptomatic transmission, a sufficiently wide and deep contact-tracing network can prevent epidemic spread through an infinite population. In order to make this point rigorously, we introduced a solvable branching-process model for recursive contact tracing at any given tracing depth.

Despite their appealing analytical tractability, such branching-process models are something of an idealization: they capture the essential features of epidemic spread in the dilute limit of an infinite population. While this ought to yield a good approximation for early-time epidemic spread in large, well-mixed populations such as cities, it cannot capture the finite-size effects that dictate the late-time behaviour of epidemics in small, local populations. Here, we address this shortcoming by developing a Reed-Frost-type model for recursive contact tracing that is applicable to finite populations and arbitrary contact networks, beyond the Bethe lattice or Cayley tree structure implicit in branching-process models.

The paper is organized as follows. In Section II, we introduce a network Reed-Frost model that incorporates both recursive contact tracing and asymptomatic transmission. We first study the model on a fully connected, or complete, graph and compare it with the branching-process model introduced in previous work~\cite{DHI}, which it recovers in the limit of infinite population size. We verify numerically that on the complete graph, the critical finite-size scaling of outbreak size and duration lies in the universality class of mean-field percolation~\cite{ben2004size,BK04,ben2012scaling}. This is consistent with our earlier findings in the branching process limit~\cite{DHI}. In Section III, we perform a numerical study of the Reed-Frost model with recursive contact tracing on the square lattice. We present evidence for a contact-tracing phase transition on the square lattice as $N \to \infty$, whose finite-size scaling near the critical line lies in the universality class of two-dimensional percolation~\cite{Cardy_1985,Tome_2010_SIR,ARGOLO_2011_2D,de_Souza_2011_SIR}. As for the Bethe lattice~\cite{DHI}, we find that the critical line connects smoothly to the singular point of perfect contact tracing and purely asymptomatic transmission, where universal behaviour gives way to a discontinuous phase transition. This demonstrates that the unusual phenomenology of the contact-tracing phase transition identified in Ref.~\cite{DHI} is independent of the network under consideration. Finally, in Section IV, we model the practically important question of how far contact tracing can control the size of epidemic outbreaks in regimes where epidemic spread is not contained. 

\section{Reed-Frost model with contact tracing}
\label{sec:SecII}

The Reed-Frost model is a discrete-time, stochastic, compartmental model that describes epidemic spread in a homogeneous population of size $N$. In each generation $n=0,1,2,\ldots$, it is assumed that there are $S_n$ susceptible individuals, $I_n$ infectious individuals and $R_n$ recovered individuals, with $S_n + I_n + R_n = N$. In a given generation, each infectious individual has a probability $q$ of infectious contact with each susceptible individual. All possible infectious contacts are allowed and modelled as independent Bernoulli trials. Given a realization of the model at time $n$, the number of infections at time $n+1$ is drawn from the probability distribution

\begin{align}\label{standard}
    I_{n+1} \sim \mathrm{Bin}(S_n,1-(1-q)^{I_n})
\end{align}
We assume that infectious individuals recover after one generation, i.e. $R_{n+1}=I_n$ and $S_{n+1}=S_{n}-I_{n+1}$.

For large population sizes $N$ and initial infections $I_0 = \mathcal{O}(1)$, the early-time dynamics of the Reed-Frost model with contact probability $q=R_0/N$ recovers a branching process model with basic reproduction number $R_0$~\cite{BARBOUR2004173}. For finite $N$, this implies an approximate critical point at $q=1/N$, which tends to the branching-process critical point $R_0=1$ in the scaling limit $qN=1, \, N\to \infty$. At late times, the finite population size $N$ cuts off the growth of the branching process, due to depletion of the susceptible population $S_n$. This effect is most dramatic near criticality, where $N$ becomes the only scale in the problem and various scaling laws related to the mean-field percolation transition emerge~\cite{ben2012scaling,BK04}. 

For the standard Reed-Frost model, Eq. \eqref{standard}, the connection with percolation can be made precise by embedding the model within an Erd\H{o}s–R{\'e}nyi random graph~\cite{barbour1990epidemics,erdHos1960evolution}, which has a natural interpretation in terms of bond percolation. For the Reed-Frost models with contact tracing that we introduce below, a straightforward interpretation in terms of bond percolation is lost, because contact tracing introduces correlations between non-adjacent bonds. Nevertheless, we find numerical evidence for finite-size scaling in the percolation universality class.

\subsection{Reed-Frost model with asymptomatic infections}

Before introducing the complexities of contact tracing, we first consider a modification of the standard Reed-Frost model that is necessary for modelling a disease like COVID-19 with substantial non-symptomatic transmission, and introduce separate compartments for symptomatic and asymptomatic infections. Let $I^A_n$ and $I^S_n$ denote the populations in each compartment in generation $n$. These populations are quantified by a probability $\theta$ of asymptomatic infection, with $I^A_n$ drawn from a binomial distribution
\begin{equation}
I^A_n \sim \mathrm{Bin}(I_n,\theta) 
\end{equation}
at each generation, and $I^S_n$ given by $I^S_n = I_n - I^A_n$. To be fully general, we assume that in each generation, asymptomatic infections and symptomatic infections have distinct probabilities of infectious contact, respectively $q_A$ and $q_S$. Thus, given a realization at time step $n$, the total number of infections in generation $n+1$ is drawn from the binomial distribution
\begin{align}
    I_{n+1} \sim \mathrm{Bin}(S_n,1-(1-q_S)^{I^S_{n}}(1-q_A)^{I^{A}_n}).
\end{align}
Infectious individuals are assumed to recover as in the standard Reed-Frost model defined around Eq. \eqref{standard}. In the scaling limit $q_A N = R_0,\, q_S N = R_S, \, N\to \infty$, this recovers the branching-process model considered in previous work~\cite{DHI} (without contact tracing).

\subsection{Reed-Frost model with contact tracing}

Let us now consider introducing contact tracing in the Reed-Frost model. An apparent difficulty is that the model has no notion of network structure. On the one hand, the dynamics of the model is simple enough that it extends easily to arbitrary network connectivities. On the other hand, the probabilistic nature of the contact infection means that the question of ``who infected whom'', that must be answered to trace contacts, becomes a Bayesian inference problem. Though this inference problem is complicated in general~\cite{teunis2013infectious}, it is solvable in the Reed-Frost model. (This difficulty does not arise in the branching process limit~\cite{DHI}, in which every infection can be traced to a unique source.)

We first define the standard Reed-Frost model on a graph. Therefore consider a graph $G$ with $N$ vertices, $i=1,2,\ldots,N$. Write $\langle i j \rangle$ if the vertices $i$ and $j$ are connected by an edge of $G$. Let $\{\hat{S}_n, \, \hat{I}_n,\, \hat{R}_n\}$ denote the sets of susceptible, infected and recovered vertices at each time step, and let
\begin{equation}
S_n^j = \begin{cases} 1 & j \in \hat{S}_n \\ 0 & j \notin \hat{S}_n \end{cases}, \quad S_n = |\hat{S}_n| = \sum_{j=1}^N S_n^j,
\end{equation}
similarly for compartments $I$ and $R$. The probability that a vertex $i$, susceptible in generation $n$, becomes infected in generation $n+1$, is determined by its infected graph neighbours at time $n$, through the Bernoulli trial
\begin{equation}
I_{n+1}^j \sim \mathrm{Bernoulli}(1-p^j_n),\quad p^j_n=(1-q)^{\sum_{\langle ij \rangle } I^i_n}.
\end{equation}
As above, we require that $S_n+I_n+R_n=N$ and $S_{n+1} = S_n-I_n$.

Let us now suppose that some subset of the population $C \subset \{1,2,\ldots,N\}$ is on a contact-tracing network. In order to trace contacts, we need to know the probability that given an infection $j \in C$ on the network at time $n+1$, a neighbouring infected vertex, $\langle i j\rangle$, infected $j$. By Bayes' rule, this is given by
\begin{equation}
\mathbb{P}(i \to j | j \, \mathrm{inf}) = \frac{\mathbb{P}(i \to j)}{\mathbb{P} (j \, \mathrm{inf})} = \frac{q}{1-p^j_n}.
\end{equation}
Edges are now added to the contact-tracing graph according to the following algorithm:
\begin{itemize}
\item[$1.$] at each new time-step $n+1$, check if there are any new infections $j$ on the contact network $C$.
\item[$2.$] for each neighbouring infection $i$ from the previous time step, $\langle i j \rangle$ with $I_{n}^i=1$, check if $i \in C$.
\item[$3.$] if both $i,j \in C$ and $I_{n}^i = I_{n+1}^j = 1$, add the edge $\langle i j\rangle$ to $C$ with probability $\frac{q}{1-p^j_n}$.
\end{itemize}

We defer a discussion of how traced contacts are isolated to the next section.

\subsection{Reed-Frost model with asymptomatic infections and contact tracing}

Finally, we define a Reed-Frost model on an arbitrary graph $G$ with both asymptomatic infections and contact tracing. This generalizes the branching-process model introduced in previous work to finite populations and arbitrary network connectivity. In the spirit of that branching-process model, let us declare at the outset that some subset $C \subset \{1,2,\ldots,N\}$ of the population is on a contact-tracing network, while another subset $A \subset \{1,2,\ldots,N\}$ of the population will not show symptoms if infected. Then any given infected individual belongs to one of four categories: on the contact-tracing network and asymptomatic (CA), on the contact-tracing network and symptomatic (CS), off the contact-tracing network and asymptomatic (NA) and off the contact-tracing network and symptomatic (NS).

As above, we assume that infectious individuals with symptoms and without symptoms have distinct probabilities, $q_S$ and $q_A$ respectively, of infectious contact with their susceptible neighbours. Thus the probability that a node $j$ that is susceptible at time $n$ becomes infected at time $n+1$ is distributed as
\begin{align}
\nonumber I_{n+1}^{j} &\sim \mathrm{Bernoulli}(1-p^j_n), \\
p^j_n &= (1-q_S)^{\sum_{\langle ij \rangle}^{i \in S} I^i_n}(1-q_A)^{\sum_{\langle i j \rangle}^{i \in A} I^i_n}
\end{align}

The contact-tracing graph is constructed as in the previous section, but inferring ``who infected whom'' is slightly more complicated, and the probabilities for adding edges must be modified accordingly:
\begin{itemize}
    \item[$3'.$] if both $i,j \in C$ and $I_{n}^i = I_{n+1}^j = 1$, add the edge $\langle i j\rangle$ to $C$ with probabilities:
    \begin{equation}
    \mathbb{P}(i \to j | j \, \mathrm{inf}) =         \begin{cases}
        \frac{q_A}{1-p^j_n} & i\in A \\
        \frac{q_S}{1-p^j_n} & i \notin A
        \end{cases}
    \end{equation}
\end{itemize}

\begin{figure}[t]
    \centering
    \includegraphics[trim=1cm 12cm 0cm 0cm,width=0.9\linewidth]{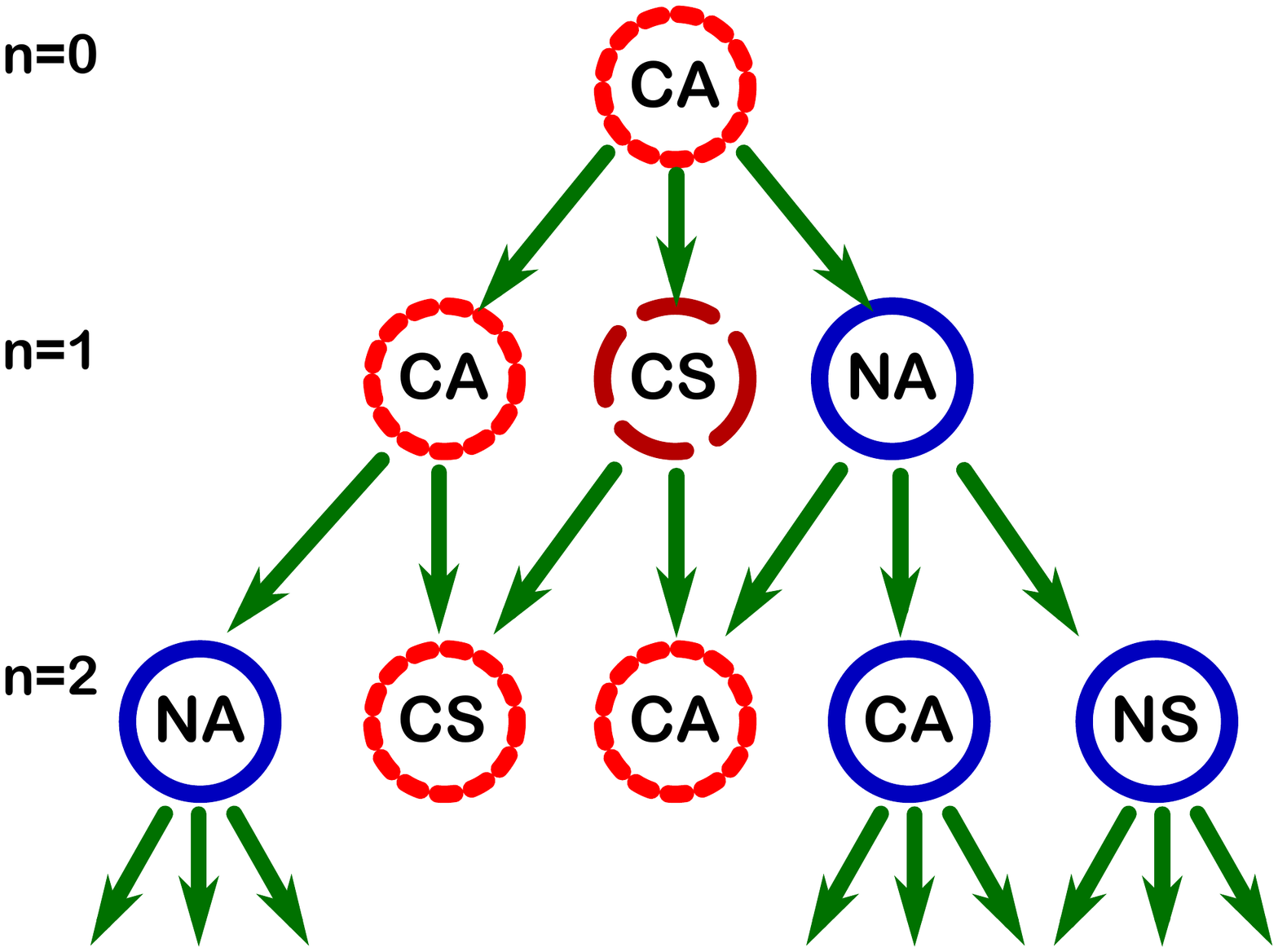}
    \caption{Sample time-evolution in a Reed-Frost model with contact tracing. The dashed and dotted circles form a contact-network-connected cluster, which is detected in generation $n=1$, when an individual on the network shows symptoms. By generation $n=2$, subsequent infections on the contact network have been traced and isolated, preventing further transmission from this cluster on the contact network, though transmission continues freely off the contact network. Observe that due to the possibility of an infected node having more than one infectious parent, the contact-network-connected cluster shown here contains a cycle. This possibility does not arise in branching-process models.}
    \label{rf_tree}
\end{figure}

We model isolation of traced contacts along the same lines as in previous work~\cite{DHI}, and briefly summarize that procedure here. In each time step $n$, the contact network is ``triggered'' whenever a CS individual is encountered. The real-world picture to have in mind, say in terms of a digital contact-tracing app, is as follows: upon identification of a newly infected CS individual $j$ in generation $n$, the app will locate all their contacts recursively, going both backwards and forwards in time until all the individuals in the $C$-connected component of $j$ up to generation $n$ are identified. These individuals are assumed to be traced, tested and isolated by generation $n+1$. This means that they cannot give rise to any infections on the contact network in generation $n+1$, although the possibility remains of non-symptomatic infection off the contact network; see Fig. \ref{rf_tree} for a visualization. One important difference compared to the branching-process model is that for the Reed-Frost model on a generic graph, it is possible that an infection $j \in C$ is traced to multiple infectious sources. This means that the contact-network-connected component of $j$ is no longer a tree, but one can still trace and isolate such connected clusters as in the branching-process model. 

We emphasize that although this model is idealized, it is sufficiently expressive to account for various inefficiencies that arise in practice. For example, imperfect contact tracing can be captured by an effective reduction in the rate of contact network coverage $\phi$, while imperfect isolation after showing symptoms is captured by an effective increase in the parameter $q_S$.

\subsection{Numerical simulations on complete graphs}

We now simulate the Reed-Frost model with recursive contact tracing by random sampling. The network $G$ is assumed to be the complete graph on $N$ vertices, which is the network structure implicit in the standard Reed-Frost model. At each initialization, we assign points in $G$ randomly to the sets $A$ and $C$, independently and with probabilities $\theta$ and $\phi$ respectively. In the scaling limit $q_AN = R_0, \, q_SN = R_S, N \to \infty$, this recovers the branching-process model studied in previous work~\cite{DHI}, whose exact phase diagram is known.

\begin{figure}
    \centering
    \includegraphics[trim=1cm 3cm 2cm 1cm,width=0.95\linewidth]{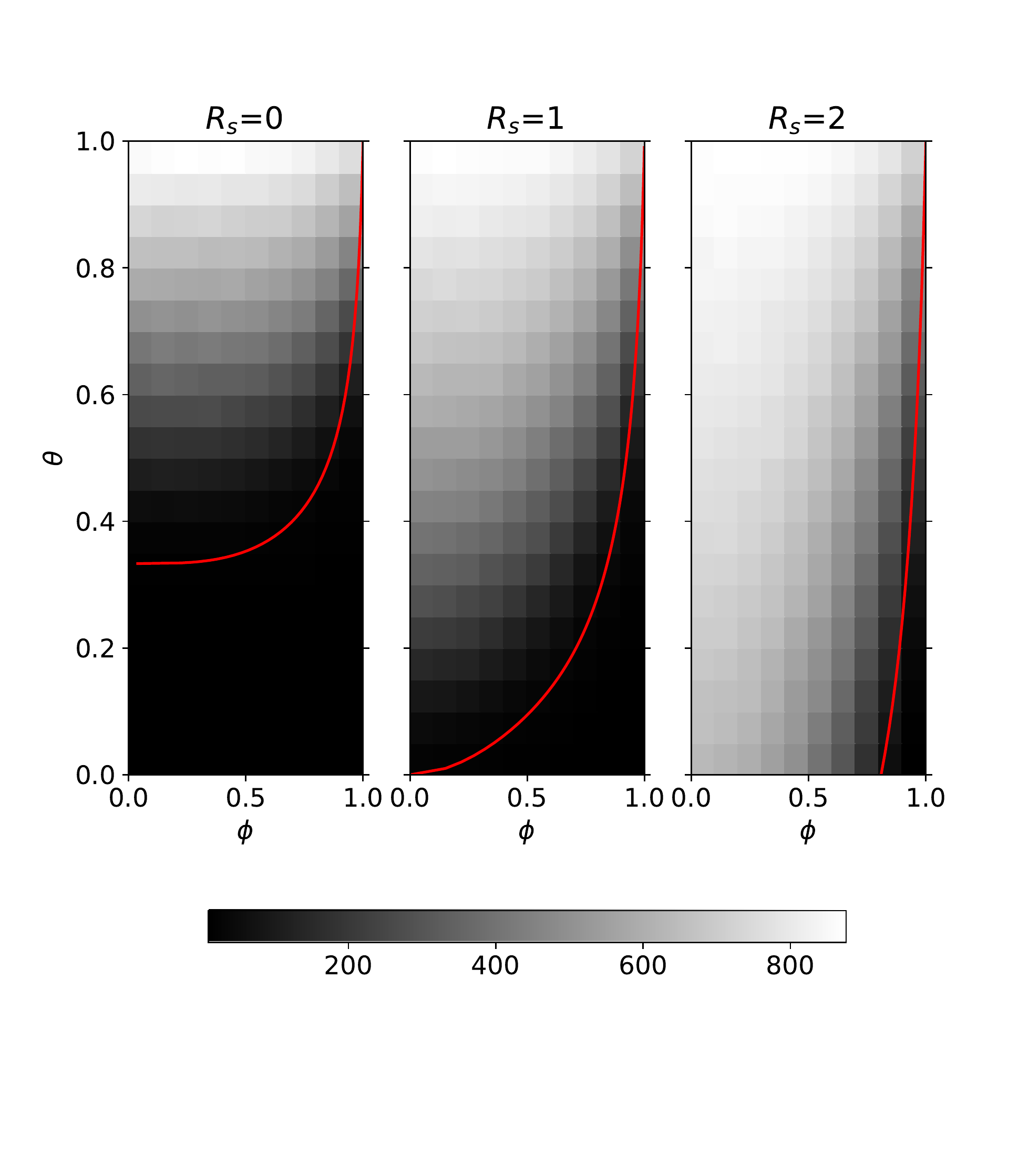}
    \caption{Average size of outbreaks for $R_0=3$ and different values $R_S=0,\,1,\,2$ in a population of size $N=1000$, averaged over 5000 realizations per point. The red curve corresponds to the critical line separating the epidemic phase from the immune phase in the corresponding $N=\infty$ branching process. The finite system size serves to increase the area of the black ``immune'' region, as discussed in the main text.}
    \label{phdiag}
\end{figure}

We initialize the model with one infected node, $I_0=1$, and consider the parameter values $q_AN=R_0=3, \, q_SN=R_S=0,1,2$ (It has been estimated that for COVID-19, $R_0=3$ is a reasonable approximation~\cite{ferretti2020quantifying} and generically, we expect $R_0\geq R_S$.). In Fig. \ref{phdiag}, we plot the outbreak size for these parameters for a population of size $N=1000$, averaged over 5000 realizations per data point. For comparison, we plot the critical line of the corresponding branching process at $N=\infty$. In a finite population, the susceptible population is depleted over time, reducing the effective reproduction number at late times. The finite population size similarly cuts off the size of epidemic outbreaks, which would grow without bound in the corresponding branching process. Both effects tend to increase the area of the subcritical region, which is consistent with what is shown in Fig. \ref{phdiag}.

We now turn to the distribution of epidemic outbreaks. As is clear from the branching process approximation, the distribution of outbreak size and duration in the standard Reed-Frost model depends sensitively on the value of $qN$. Deep in the ``epidemic'' regime $qN>1$, a bi-modal distribution function $f(I)$ for epidemic sizes is expected, where the random variable
\begin{equation}
I = \sum_{n=0}^\infty I_n
\end{equation}
denotes the total size of the epidemic. Both small and large outbreaks are possible, with large outbreaks distributed approximately normally about their mean for large $N$~\cite{BARBOUR2004173}. In this regime, the mean size of epidemics is proportional to $N$. Near criticality, $qN \approx 1$, the distribution function of epidemic sizes exhibits power-law scaling~\cite{ben2004size,ben2012scaling}, which we discuss in more detail below. Deep in the ``immune'' regime $qN < 1$, outbreaks are expected to be small and short-lived, since they sample the $\mathcal{O}(N^0)$ mean cluster size of subcritical percolation.

\begin{figure}
    \centering
    \includegraphics[trim=0.5cm 0cm 0.5cm 0cm,width=\linewidth]{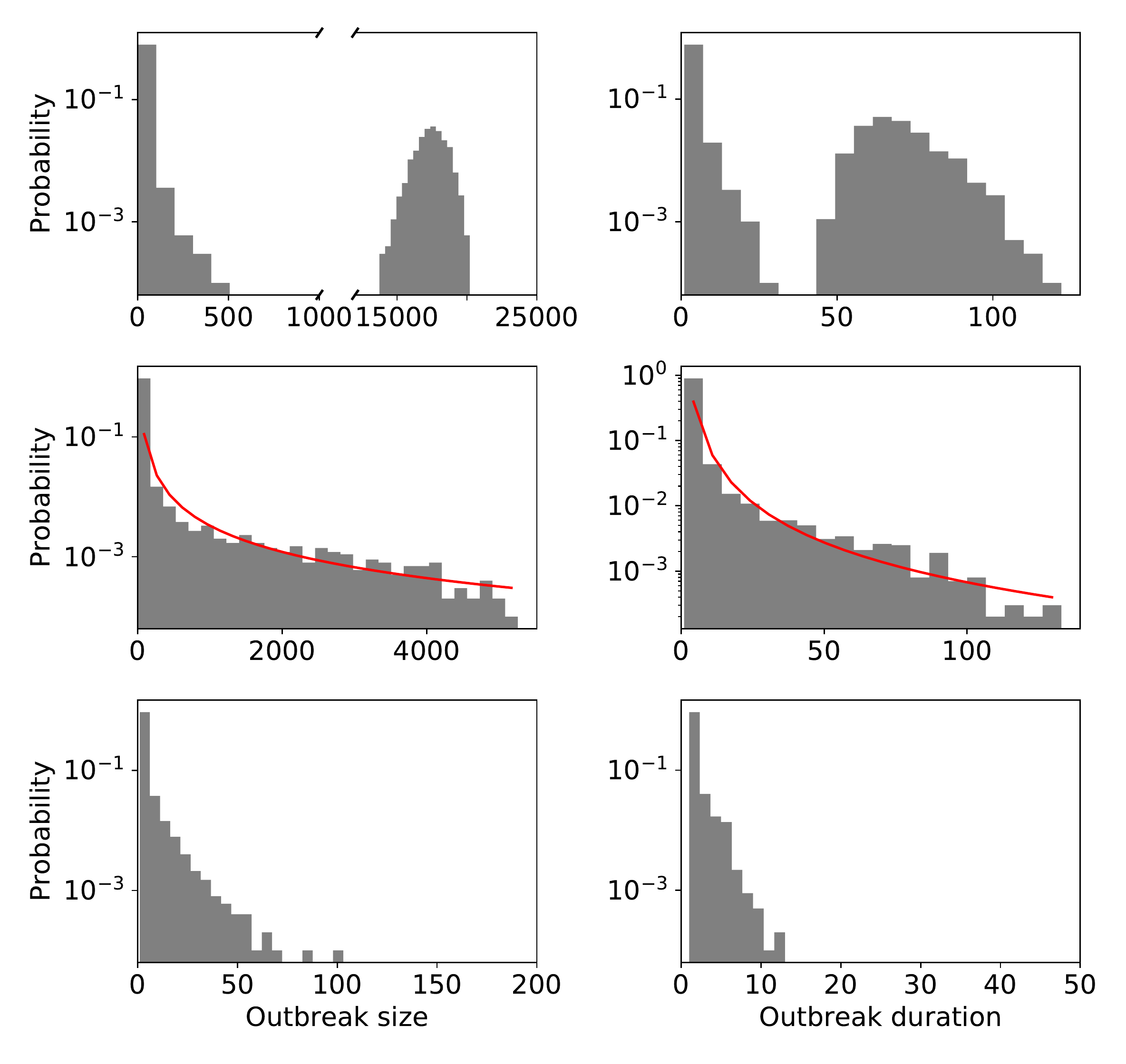}
    \caption{Histograms of outbreak size (left) and duration (right) on the complete graph, with parameters $R_0=3,R_S=0$ in a population of size $N=50000$ and $10000$ realizations for each row. Deep in the epidemic phase (top row, $\phi=0.7,\theta=0.5$), we observe bi-modal distributions with many large and many small outbreaks. Near criticality (middle row, $\phi=0.7,\theta=0.4$) there are power-law tails of large and long outbreaks. The fitted power laws (red curves) have exponents $-1.45 \pm 0.16$ for outbreak size and $-2.02\pm 0.17$ for outbreak duration, which are both consistent with known values for the standard Reed-Frost model~\cite{ben2004size} (respectively $-3/2$ and $-2$). Moving into the immune phase (bottom row, $\phi=0.7,\theta=0.2$), these long tails become exponentially suppressed, and no large epidemics are observed. The qualitative behaviour in all three regimes is as one would expect for the standard Reed-Frost model, but arises here from the competition between contact tracing and asymptomatic transmission.}
    \label{dist}
\end{figure}

We now check that these expectations are met in the Reed-Frost model with recursive contact tracing. Figure \ref{dist} shows numerical probability distribution functions for the outbreak size $I$ and duration $t = \sum_{n=0}^\infty (1-\delta_{I_n0})$, for three different pairs of parameter values $(\phi,\,\theta)$ and $R_0=3,\,R_S=0$, in a population of size $N=50,000$ with 10,000 realizations for each set of parameters.

We next verify that the critical finite-size scaling behaviour of the contact-tracing phase transition on the complete graph matches the power-law distributions observed in the standard Reed-Frost model. Let us first summarize how these scaling laws can be understood from percolation theory~\cite{ben2012scaling}. At criticality, the distribution function for outbreak sizes is given by the mean-field percolation result~\cite{Grimmett_1999}, $f(I)\sim I^{-3/2}$. Introducing the maximum outbreak size $I^*$ as a cutoff in this expression implies that the mean outbreak size $\mathcal{I}$ is related to the maximum outbreak size as $\mathcal{I}^2 = I^*$. One might expect that the maximum outbreak size $I^* \propto N$, as occurs on the epidemic side of the transition. In fact this is not the case, due to the depletion of susceptible individuals at late times~\cite{ben2004size}. The correct scaling behaviour can be obtained in the percolation language from the size of giant components in the critical Erd\H{o}s-R{\'e}nyi problem~\cite{erdHos1960evolution,barbour1990epidemics,BK04,ben2012scaling}, which yields $I^* \sim N^{2/3}$. From this it follows that the mean outbreak size scales as $\mathcal{I}\sim N^{1/3}$. By exponential growth of the underlying branching process, one can further argue that the mean outbreak duration $\langle t \rangle \sim \log{\mathcal{I}} \sim c\log{N}$, for some constant $c$. This is consistent with known scaling forms for the typical diameter of near-critical Erd\H{o}s-R{\'e}nyi graphs~\cite{CHUNG2001257}, but more detailed arguments are needed to fix the prefactor $c=1/3$~\cite{ben2012scaling}. Finally, we note that the distribution function for epidemic durations $g(t)\sim t^{-2}$~\cite{ben2004size,ben2012scaling}.

In Figure \ref{dist} (middle row) we compare the distribution functions of outbreak sizes and duration near the critical line of the contact-tracing phase transition with that of a standard Reed-Frost model with $q=1/N$, for population size $N=50,000$ and parameter values $R_0=3,\,R_S=0, \, \phi=0.7,\,\theta=0.4$ in the contact-tracing model. We find that the distribution functions for both outbreak size and duration observed for critical contact-tracing on the complete graph are consistent with the critical behaviour of the standard Reed-Frost model, which supports our earlier claim~\cite{DHI} that on almost all of the critical line, the contact-tracing phase transition lies in the universality class of mean-field percolation.

\begin{figure}
    \centering
    \includegraphics[trim=0cm 0cm 0cm 0cm,width=0.95\linewidth]{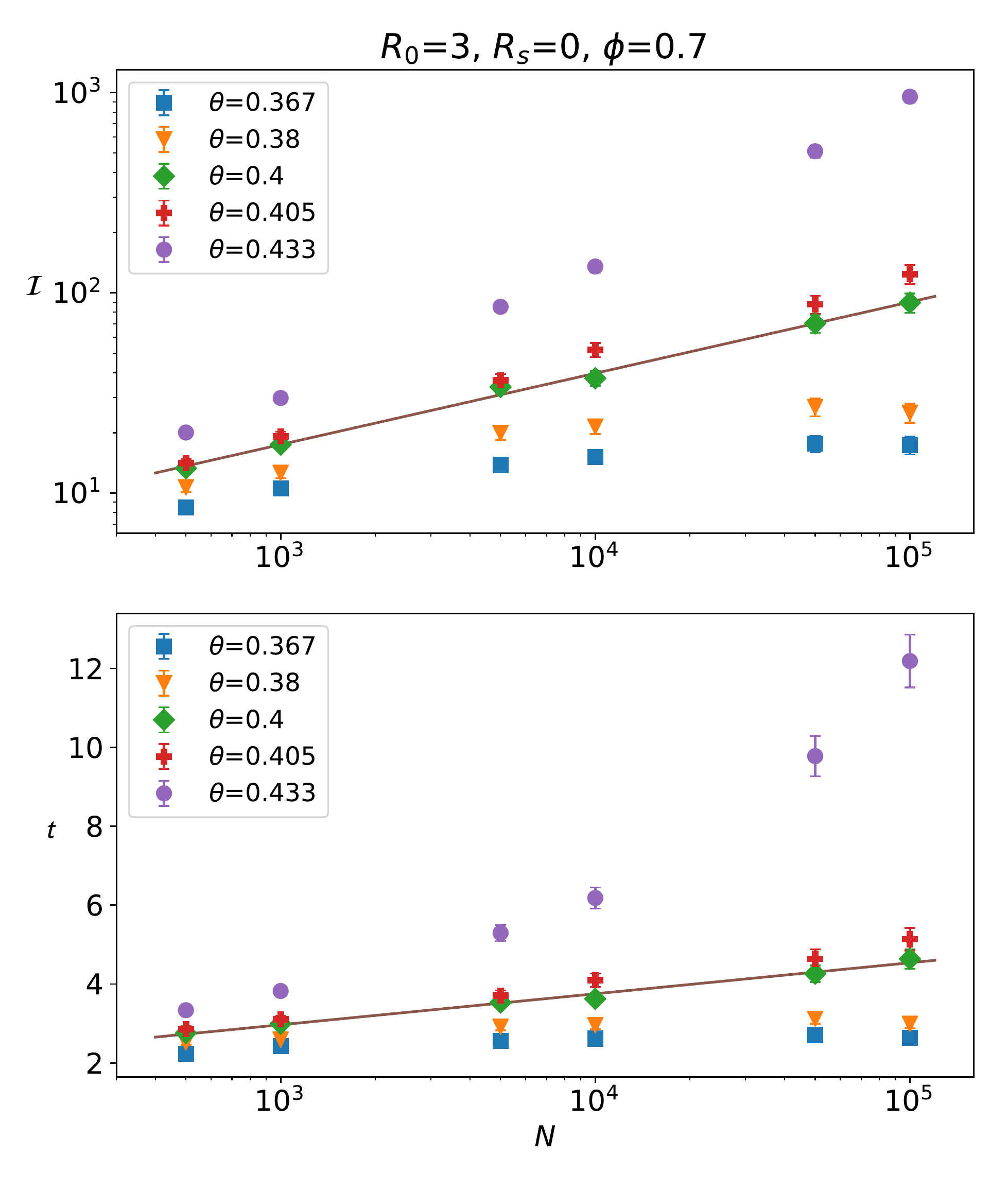}
    \caption{Top: average outbreak size $\mathcal{I}$ versus total population size on the complete graph, near the critical point $R_0=3,R_S=0,\phi_c=7,\theta_c=0.4$ of the $N=\infty$ branching process, averaged over $10000$ realizations for each $\theta$ and $N$. The slope of the solid line is $0.35\pm0.01$, which is consistent with its exact value $1/3$ in the standard Reed-Frost model. Bottom: mean outbreak duration versus total population size near the critical point. The solid line has a slope of $0.34\pm 0.02$, consistent with the standard Reed-Frost prediction $t\sim\frac{1}{3}\mathrm{log}(N)$.}
    \label{phi_size}
\end{figure}

\begin{figure}
    \centering
    \includegraphics[trim=0cm 0cm 0cm 0cm,width=0.8\linewidth]{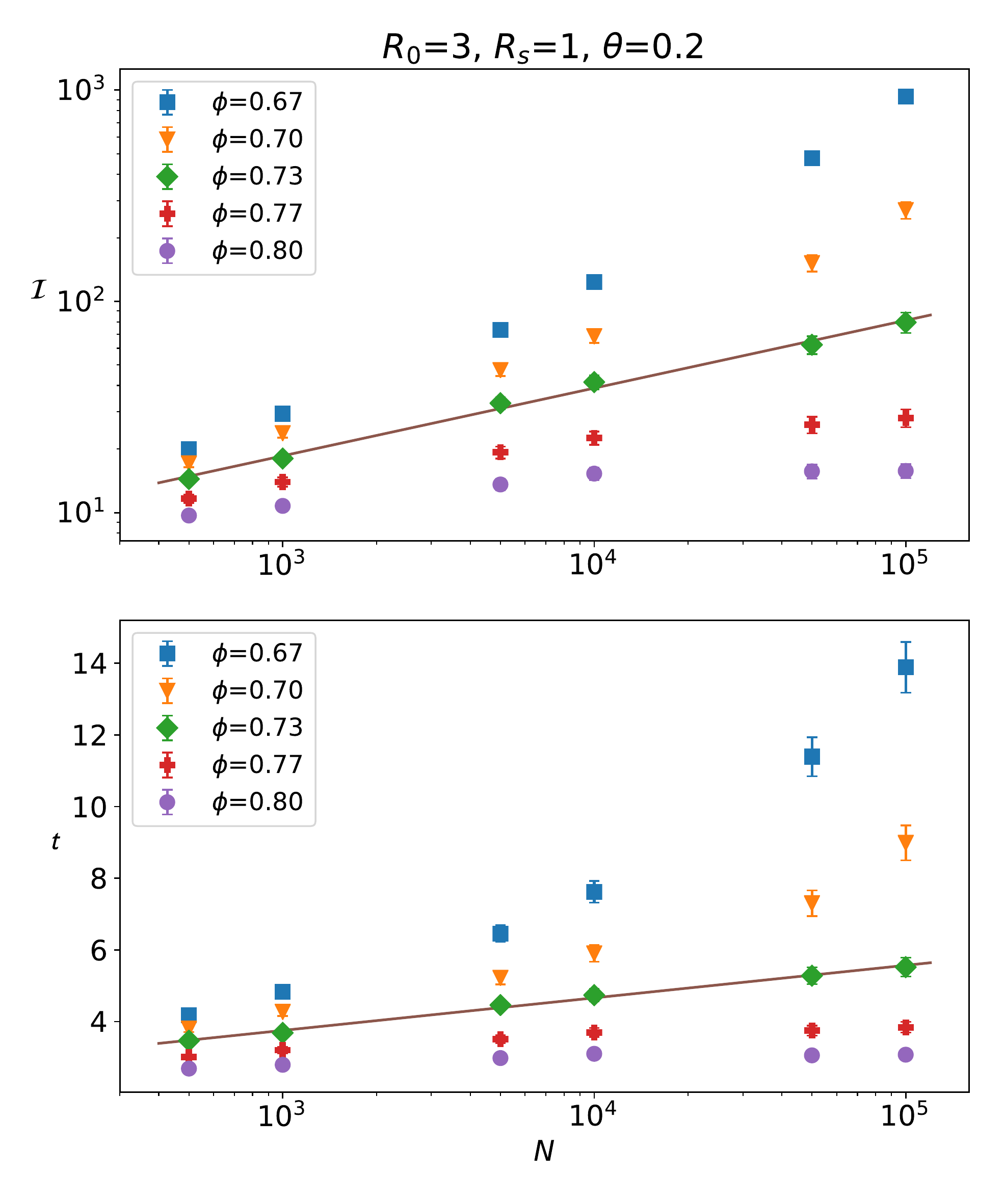}
    \caption{Top : average outbreak size $\mathcal{I}$ versus total population size on the complete graph, near the critical point $R_0=3,R_S=1,\phi_c=0.71,\theta_c=0.2$ of the $N=\infty$ branching process, averaged over $10000$ realizations for each $\phi$ and $N$. The slope of the solid line is $0.32\pm0.01$, which is consistent with its value $1/3$ in the standard Reed-Frost model. Bottom: mean outbreak duration versus total population size near the critical point. The solid line has a slope of $0.39\pm0.01$, which is close to the prediction $t\sim\frac{1}{3}\mathrm{log}(N)$ of the standard Reed-Frost model. The small discrepancy between the branching process critical point $\phi_c=0.71$ and the observed value of around $\phi_c \approx 0.73$ may be attributed to finite-size effects.}
    \label{theta_size}
\end{figure}

\begin{figure}
    \centering
    \includegraphics[trim=0cm 0cm 0cm 0cm,width=0.8\linewidth]{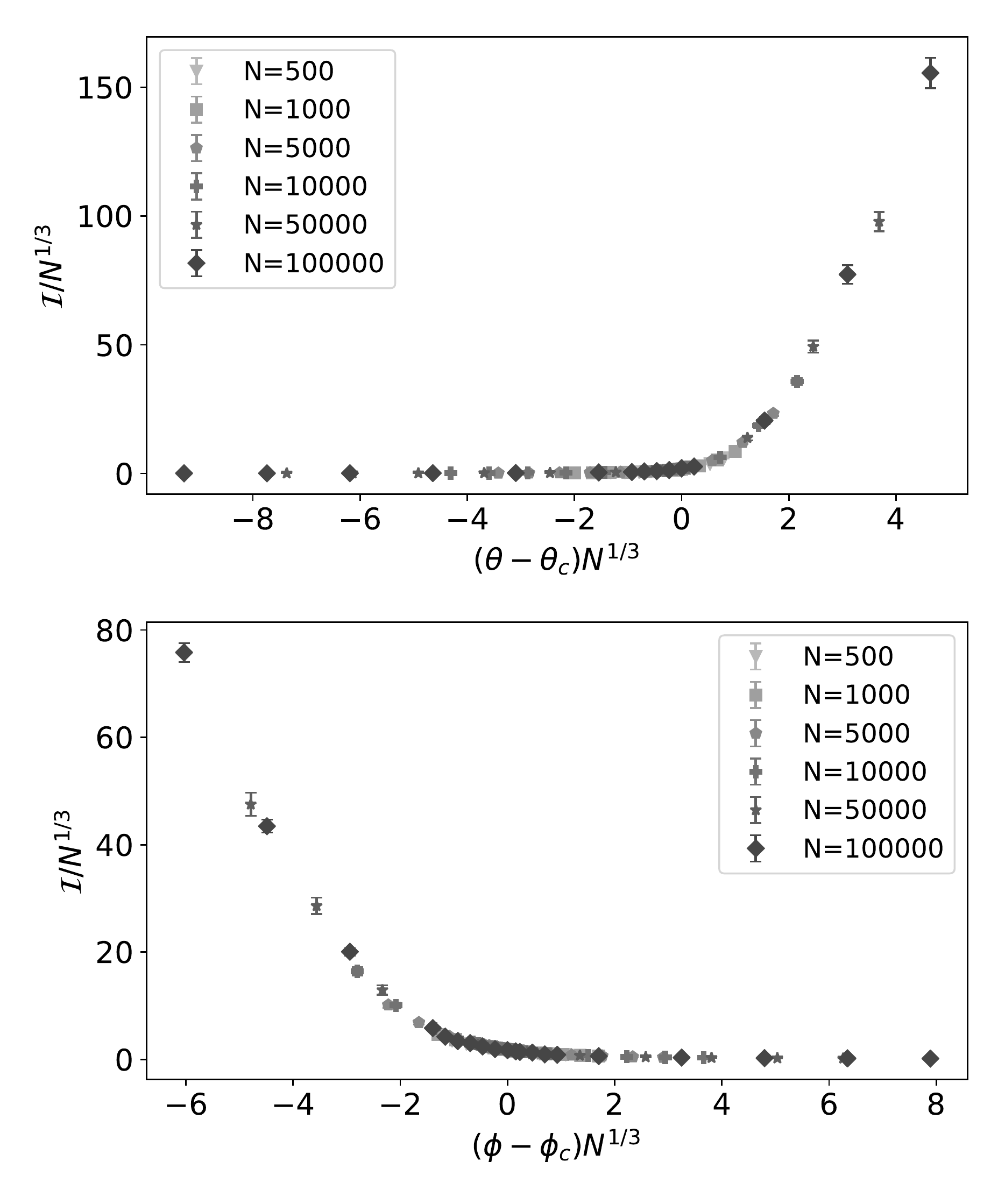}
    \caption{Finite-size scaling of the average outbreak size $\mathcal{I}$ on the complete graph, for the critical points $\theta_c=0.4$ for $R_0=3,R_S=0,\phi=0.7$ (top) and $\phi_c=0.73$ for $R_0=3,R_S=1,\theta=0.2$ (bottom) for different system sizes, averaged over 10000 realizations for each size, with the scaling exponents chosen to match the standard Reed-Frost model.}
    \label{collapse}
\end{figure}

We next address the finite-size scaling of average outbreak size and duration near criticality. To this end, we simulate system sizes $N=500,\,1000,\,5000,\,10000,\,50000,\,100000$ and study the scaling of average outbreak sizes and duration at the critical point along two directions : along the $\theta$ direction for $R_0=3,\,R_S=0,\,\phi=0.7$, and along the $\phi$ direction for $R_0=3,\,R_S=1,\,\theta=0.2$.
For each system size, we average over $10000$ realizations. Since we expect the critical points of the envelope branching process ($\theta_c=0.4,\,\phi_c=0.7$ for $R_0=3,\,R_S=0$ and $\theta_c=0.2,\,\phi_c=0.71$ for $R_0=3,\,R_S=1$) to dictate the crossovers in the finite system, up to finite-size effects), we focus on the vicinity of these points. Results for finite-size scaling along the $\theta$ direction are recorded in Fig. \ref{phi_size}, while finite-size scaling along the $\phi$ direction is considered in Fig. \ref{theta_size}. In both cases, the observed scaling is consistent with the universal behaviour of the standard Reed-Frost model discussed above. The scaling collapse of $\mathcal{I}/N^{1/3}$ plotted against $(\theta-\theta_c)N^{1/3}$ and $(\phi-\phi_c)N^{1/3}$ in Fig. \ref{collapse} also confirms that the quantities $\theta/\theta_c$ and $\phi/\phi_c$ exhibit the same finite-size scaling as the critical reproduction number, $R_0\approx 1 + \mathrm{const.}\times N^{-1/3}$, of the standard Reed-Frost model~\cite{ball_Rscaling}. All error bars plotted denote a $95\%$ confidence interval about the mean.


\section{Square Lattice}
\label{sec:SqLat}

In previous work, we argued that the contact-tracing phase transition on the Bethe lattice, away from the singular line $\phi=1$, lies in the universality class of mean-field percolation~\cite{DHI}. We presented numerical evidence that this universal behaviour was insensitive to the recursive tracing depth $n$, even in the limit $n \to \infty$. At first sight, this result is somewhat surprising as recursive contact tracing is a highly non-local process, whose non-locality only increases with $n$. One way to understand this behaviour is through the following qualitative renormalization group argument. For any finite tracing depth $n$, one can imagine coarse-graining the system into blocks with generational depth $N>n$. Then, although recursive contact tracing is effective within each block, it does not mitigate epidemic spread between blocks, and the renormalized epidemic dynamics after this blocking transformation is effectively local in time. This argument suggests mean-field critical behaviour in the entirety of the phase diagram. It becomes less plausible in the limit $n=\infty$, and breaks down completely on the singular line $\phi=1$, which exhibits a discontinuous phase transition. Nevertheless, for $\phi<1$, this qualitative argument for mean-field behaviour is consistent with the rapid numerical convergence of exact $n=\infty$ predictions from percolation theory, viewed as power series in $\phi$~\cite{DHI}.

With the aid of the model introduced in Section \ref{sec:SecII}, one can move away from the mean-field limit of contact tracing on the Bethe lattice, and consider the contact-tracing phase transition on lattices with an effective dimensionality lower than the upper critical dimension for percolation, $d_c=6$. For concreteness, let us focus on the square lattice, for which $d=2<d_c$. Previous works have established that the critical SIR model on the square lattice lies in the same universality class as two-dimensional percolation~\cite{Tome_2010_SIR,ARGOLO_2011_2D,de_Souza_2011_SIR}. A natural question is whether the qualitative renormalization group argument above extends to the square lattice, i.e. whether the non-locality of recursive contact tracing can again be removed by a blocking transformation that recovers the standard universality class. Below, we provide numerical evidence that the contact-tracing phase transition on the square lattice indeed lies in the universality class of two-dimensional percolation, with singular behaviour on the line $\phi=1$.

\subsection{Outbreak size}

We now simulate the model described in Section \ref{sec:SecII} numerically, where the graph $G$ is a square lattice with side length $L=33$ and population size $N=L^2=1089$. We assume periodic boundary conditions, which should not affect universal behaviour as $N \to \infty$, and start from a single initial infection $I_0=1$ at the centre of the lattice.

\begin{figure}
    \centering
    \includegraphics[trim=2cm 3cm 2cm 1cm,width=0.95\linewidth]{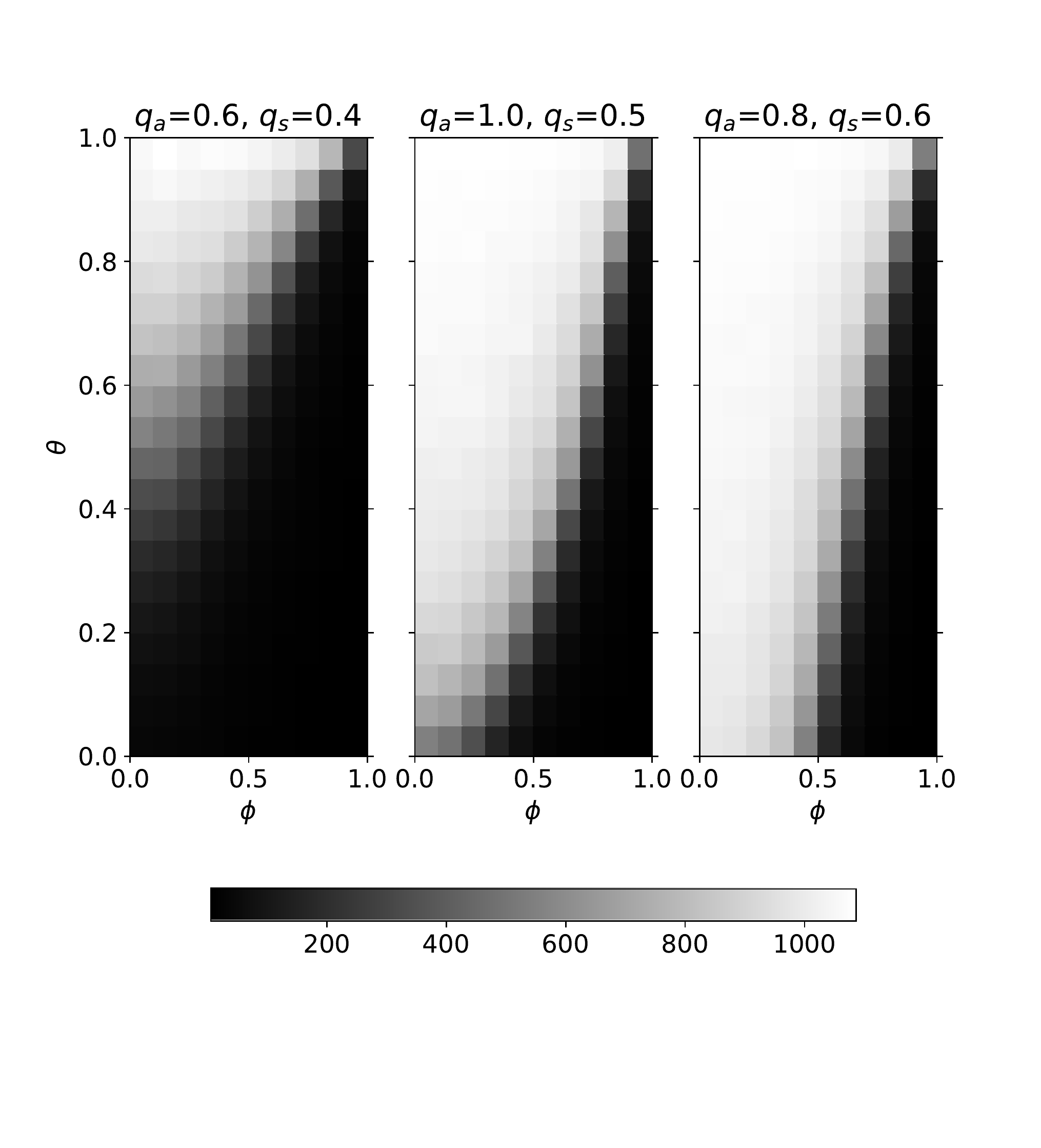}
    \caption{Average outbreak size for different asymptomatic and symptomatic contact probabilities ($q_a=0.6,q_s=0.4$ for the left, $q_a=1.0,q_s=0.5$ for the middle and $q_a=0.8,q_s=0.6$ for the right plot) for the Reed-Frost model with recursive contact tracing on a square lattice of side length $L=33$, averaged over 2000 realizations per point. The observed behaviour is qualitatively similar to that shown for the complete graph in Fig. \ref{phdiag}.}
    \label{fig:sq_phdiag}
\end{figure}

In Fig. \ref{fig:sq_phdiag}, we simulate the mean outbreak size, averaged over $2000$ realizations for three different choices of the infectious contact probabilities $\{q_A,\,q_S\}$ and plotted as a phase diagram in $\phi$ and $\theta$. We observe a blurred phase transition that is qualitatively similar to the different regimes found by varying the rate of symptomatic transmission on the complete graph, Fig. \ref{phdiag}, although for the Reed-Frost model on the square lattice there are no analytical results as $N\to\infty$ available for comparison (such results may be exist in principle but their derivation in $d=2$ is expected to be more involved~\cite{kesten1980critical} than in $d=\infty$). Notice that the ``immune phase'' again extends all the way up to the point $(\phi=1,\theta=1)$, indicating a discontinuous phase transition at this point, just as for the Bethe lattice~\cite{DHI}.

\subsection{Finite-size scaling}

\begin{figure}
    \centering
    \includegraphics[trim=0cm 0cm 0cm 0cm,width=0.8\linewidth]{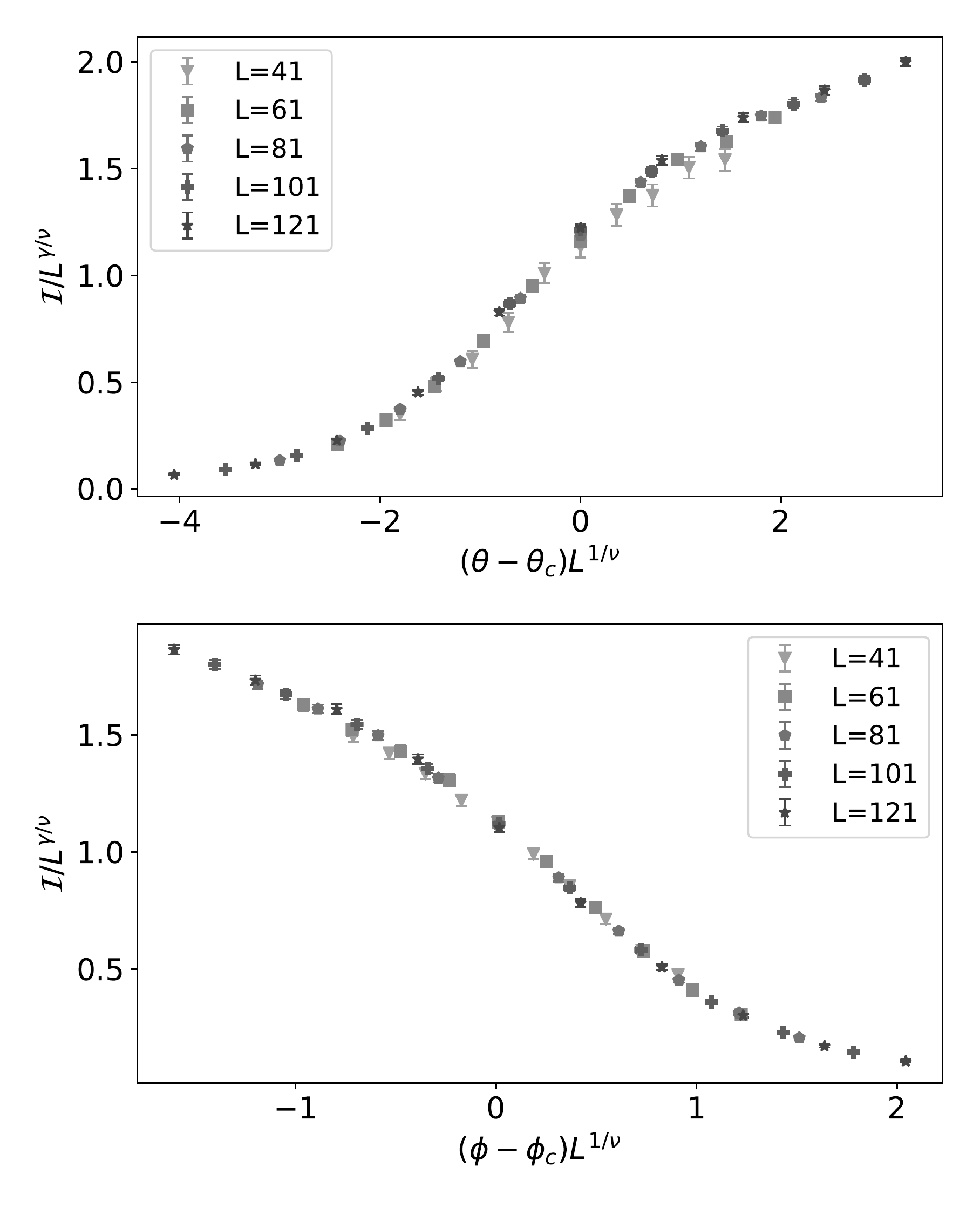}
    \caption{Finite-size scaling of the mean outbreak size $\mathcal{I}$ on the square lattice with infectious contact probabilities $q_A=1, \,q_S=0.5$, and parameter values $\theta_c=0.31,\,\phi=0.5$ (top) and $\phi_c=0.644,\,\theta=0.5$ (bottom), averaged over $5000-10000$ realizations for each $L$. Critical exponents are set to two-dimensional percolation predictions $\gamma/\nu=43/24\approx 1.792$ and $\nu=4/3\approx 1.333$ and show good scaling collapse.}
    \label{fig:sq_I_collapse}
\end{figure}

\begin{figure}
    \centering
    \includegraphics[trim=0cm 0cm 0cm 0cm,width=0.9\linewidth]{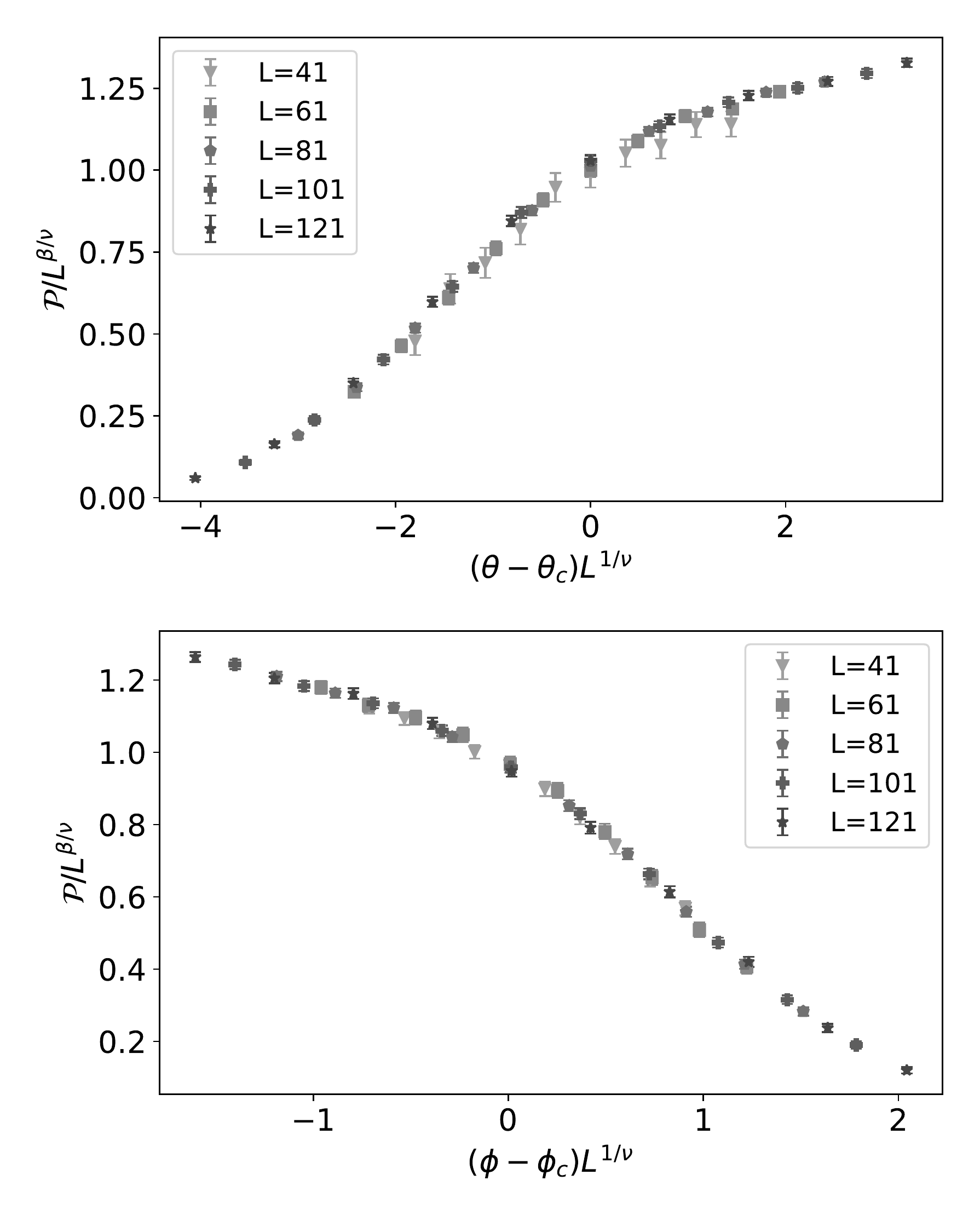}
    \caption{Finite-size scaling of the percolation probability $P$ on the square lattice with infectious contact probabilities $q_A=1, \,q_S=0.5$, and parameter values $\theta_c=0.31, \, \phi=0.5$ (top) and $\phi_c=0.644, \, \theta=0.5$ (bottom), averaged over $5000-10000$ realizations for each $L$. Critical exponents are set to two-dimensional percolation predictions $\beta/\nu=5/48 \approx 0.1048$ and $\nu=4/3\approx 1.333$ and show good scaling collapse.}
    \label{fig:sq_P_collapse}
\end{figure}

We obtain the critical exponents for the contact-tracing phase transition on the square lattice by studying the approach to the critical line from two orthogonal directions in the $(\phi,\theta)$ plane, as above. For concreteness, we fix $q_a=1.0, q_s=0.5$, as in the middle plot of Fig. \ref{fig:sq_phdiag} and study the finite-size scaling behaviour of epidemic outbreaks along the lines $\theta=0.5$ and $\phi=0.5$, respectively. Specifically, we focus on the mean outbreak size $\mathcal{I}$ and the probability $P$ that an outbreak reaches the boundary of the square lattice, which as $N \to \infty$ tends to the probability that an initial infection belongs to the infinite percolating cluster and thus defines an order parameter for the percolation phase transition. The scaling forms for $\mathcal{I}$ and $P$ can be written~\cite{de_Souza_2011_SIR} as
\begin{align}
    \mathcal{I}&=L^{\gamma/\nu}\bar{I}(L^{1/\nu}\epsilon)\\
    P&=L^{-\beta/\nu}\bar{P}(L^{1/\nu}\epsilon)
\end{align}
where $\bar{I}$ and $\bar{P}$ are the appropriate scaling functions, and $\epsilon=\phi-\phi_c$ or $\epsilon=\theta-\theta_c$. Through the scaling for $\mathcal{I}$ in Figure~\ref{fig:sq_I_collapse} and for $P$ in Figure~\ref{fig:sq_P_collapse}, we observe a good collapse to the critical exponents $\gamma/\nu= 43/24 \approx 1.792$, $\beta/\nu=5/48 \approx 0.1048$ and $\nu=4/3 \approx 1.333$ of two-dimensional percolation.


\section{Application to the epidemic phase}
\label{sec:SecIV}
As a practically relevant application of our model, we now ask how far recursive contact tracing improves epidemic control in regimes where epidemic spread is uncontrolled, i.e. on the epidemic side of the contact-tracing phase transition. Strictly speaking, questions about scaling in the epidemic phase are beyond the branching-process model studied in previous work~\cite{DHI}, which by definition pertains to infinite populations and unbounded epidemics. In contrast, the Reed-Frost model with recursive contact tracing is meaningful for finite populations, and can thus be used to study the efficacy of contact tracing even in the epidemic phase.

\begin{figure}
    \centering
    \includegraphics[trim=0cm 0cm 0cm 0cm,width=0.95\linewidth]{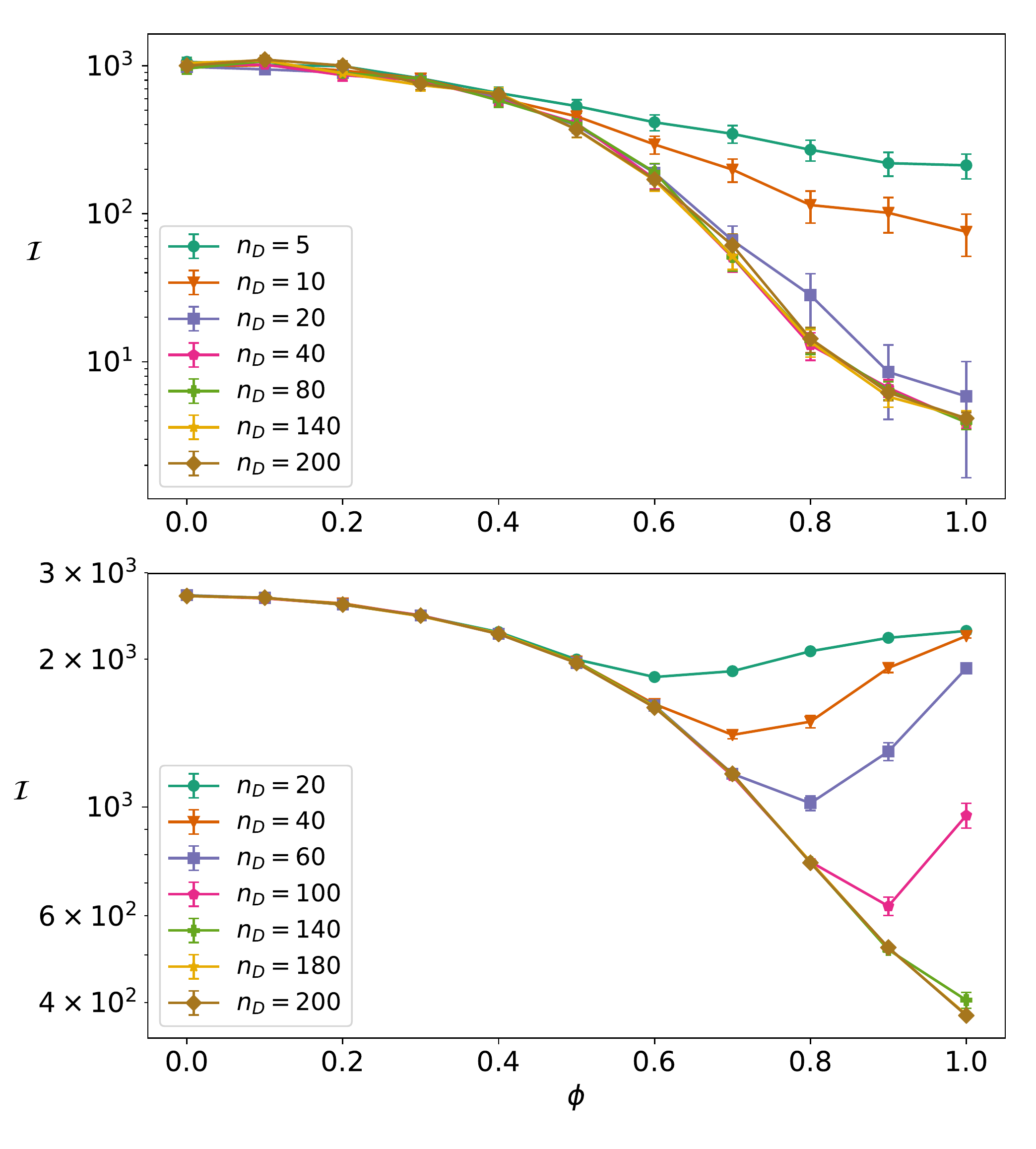}
    \caption{Mean outbreak size $\mathcal{I}$ versus $\phi$ on the complete graph, varying the maximum cluster size $n_D$ that the contact network can detect. Population size is $N=5000$ and model parameters are set to $R_0=3,\,R_S=1,\,\theta=0.2$, averaged over 1000 realizations per point. We compare the efficacy of contact tracing for controlling epidemics that start with a single infection (top) against those that start with 100 infections (bottom).}
    \label{fig:phi_nsize}
\end{figure}

We first model this problem for epidemics on the complete graph. Two examples for populations of size $N=5000$ and parameter values $R_0=3,\, R_S=1,\,\theta=0.2$, with each data point averaged over 1000 model realizations, are shown in Fig. \ref{fig:phi_nsize}. We vary both $\phi$ and the maximum allowed size of detected clusters $n_D$, which is a proxy for the recursive tracing depth. It is clear that deep in the epidemic phase (i.e. small $\phi$), contact tracing has little effect regardless of how much $n_D$ is increased. Meanwhile, for larger values of $\phi$ an immune phase is recovered for $n_D  \approx 10-100$, with larger population fractions of initial infections requiring larger values of $n_D$ before the population crosses over to a collectively immune phase. On approaching the transition from the epidemic phase, we see contact tracing starting to make a difference to the final epidemic size. For example, in Fig. \ref{fig:phi_nsize}, we see that at $\phi \approx 0.5$ the final epidemic size is down by  40\%-50\%, even though the transition to the immune phase does not take place until $\phi_c \approx 0.71$.

We also simulate the analogous problem on the square lattice, to check whether these qualitative conclusions depend on the high connectivity of the complete graph. For concreteness, we choose a square lattice of size $L=61$, averaging over 5000 realizations per point and focus on the case of one initial infection. We fix parameter values $\{q_a=1.0, q_s=0.5\}$, and sweep along the axes $\theta=0.5$ and $\phi=0.5$ of the phase diagram. The qualitative behaviour is very similar to what is observed on the complete graph, with little or no epidemic suppression deep in the epidemic phase, and convergence to epidemic control in the immune phase for values $n_D \approx 100$. The approach to the transition from the epidemic phase again shows contact tracing leading to a reduced final epidemic size, as for the complete graph, with around 15\%-20\% reduction at $\phi=0.5,\theta=0.5$ compared to the critical values $\phi_c\approx 0.64, \theta_c=0.5,$ and $\phi_c = 0.5, \theta_c\approx 0.3$ indicated by Fig. \ref{fig:sq_phdiag}. The smaller reduction compared to the complete graph can be attributed to the larger rate of asymptomatic transmission $\theta$.

\begin{figure}
    \centering
    \includegraphics[trim=0cm 0cm 0cm 0cm,width=0.95\linewidth]{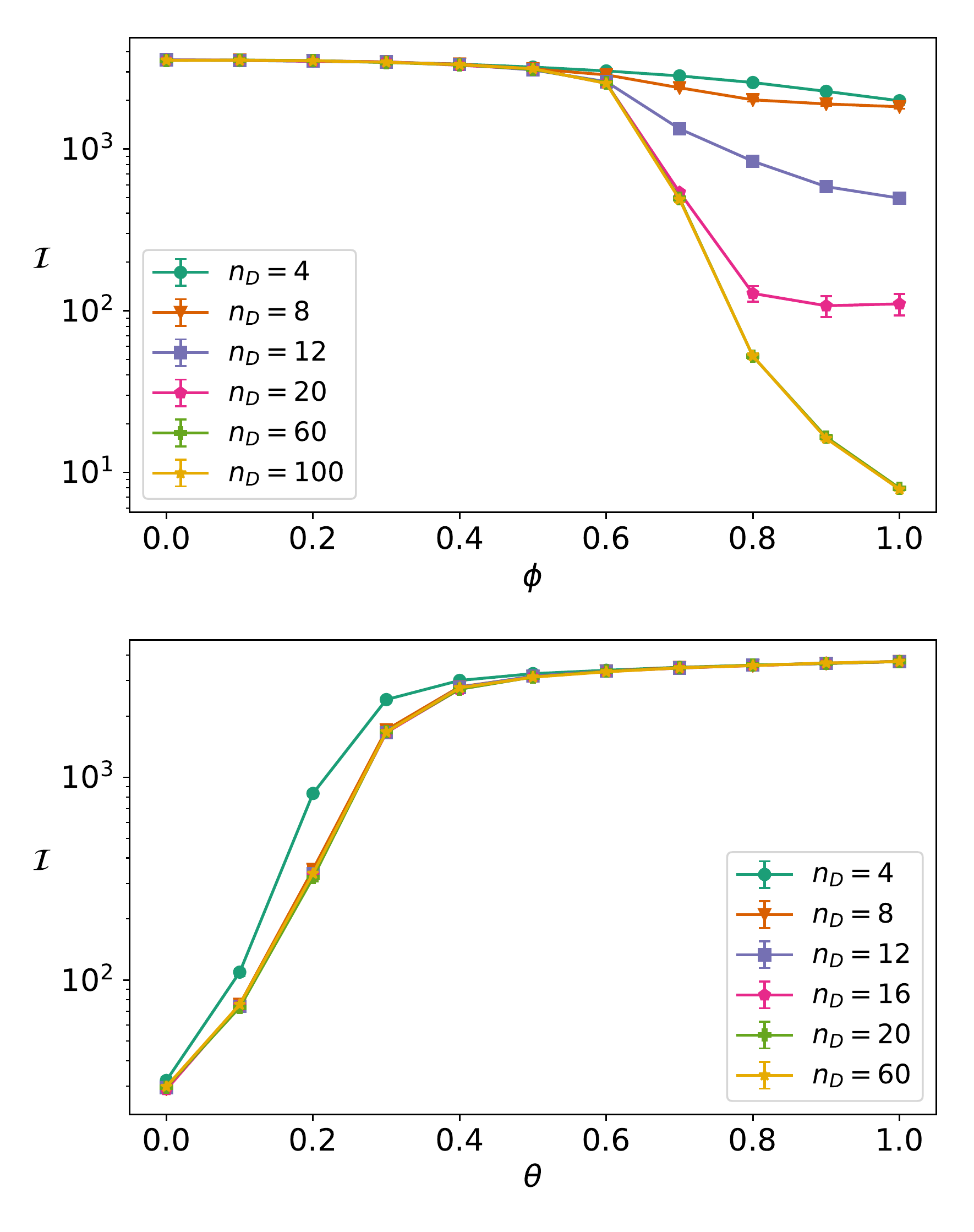}
    \caption{Mean outbreak size $\mathcal{I}$ on the square lattice, varying the maximum cluster size $n_D$ that the contact network can detect. Population size is $N=61^2=3721$ and model parameters are set to $q_A=1.0,\,q_S=0.5$, with a single initial infection $I_0=1$. We compare the efficacy of contact tracing for controlling epidemics at fixed $\theta=0.5$ (top) and fixed $\phi=0.5$ (bottom).}
    \label{fig:sq_nsize}
\end{figure}

\section{Conclusion}

We have introduced a Reed-Frost model with recursive contact tracing and asymptomatic transmission on general contact networks. This allows us to go beyond previous work~\cite{DHI} and quantify the tension between recursive contact tracing and asymptomatic spreading in finite populations. As illustrative examples, we simulated the model on the complete graph ($d \to \infty$) and on the square lattice $(d=2)$, verifying in each case that the finite-size scaling of the contact-tracing phase transition was consistent with the dimensionally appropriate percolation universality class. From the viewpoint of statistical physics, it remains to be seen whether there exist networks or models in which higher-order contact-tracing processes can drive the contact-tracing phase transition to a genuinely new universality class. The discontinuous critical point at $(\phi,\theta)=(1,1)$, observed above and in previous work~\cite{DHI} provides one tantalizing hint of a connection with recent ideas of ``explosive percolation''~\cite{SABERI20151}.  

Finally, we applied our model to the practically important question of whether contact tracing is a useful intervention when the effective reproduction number $R>1$, i.e. on the epidemic side of the epidemic-to-immune phase transition. Very far from the transition, contact tracing has little effect. However as the transition is approached, it brings down the epidemic size and for smartphone fractions of about 50\%, the reduction is significant for values of asymptomatic transmission relevant to COVID-19. We also note that the final epidemic size is related to $R$ so this is also correctly viewed as a reduction in that quantity. 

Our results suggest that when contact tracing is the only intervention in use (for example, this would be a minimally invasive way to prevent epidemic outbreaks, pending population-wide vaccination), large-scale participation is necessary for it to control epidemics entirely by itself. On the other hand, when the take-up of contact tracing is limited, our findings support the emerging consensus that it must be combined with other measures, such as vaccination, social distancing and personal protective equipment, for a ``layered protection'' that brings the effective reproduction number $R$ below one.




\section*{acknowledgments}
The authors would like to thank Sanjay Moudgalya for useful discussions. The authors are also pleased to acknowledge that the work reported on in this paper was substantially performed using the Princeton Research Computing resources at Princeton University which is consortium of groups led by the Princeton Institute for Computational Science and Engineering (PICSciE) and Office of Information Technology's Research Computing.

\bibliography{main_v2}

\end{document}